\documentclass[prl,aps,twocolumn,showpacs]{revtex4}
\usepackage[dvips]{graphicx}
\usepackage{amsmath,amssymb,float}
\usepackage{times,mathptmx}
\begin{document}
\title{Phases of the generalized two-leg spin 
ladder: A view from the SU(4) symmetry}
\author{P. Lecheminant}
\affiliation{Laboratoire de Physique Th\'eorique et
Mod\'elisation, CNRS UMR 8089,
Universit\'e de Cergy-Pontoise, 5 Mail Gay-Lussac, Neuville sur Oise,
95031 Cergy-Pontoise Cedex, France}
\author{K. Totsuka}
\affiliation{Yukawa Institute for Theoretical Physics, 
Kyoto University, Kyoto 606-8502, Japan}
\begin{abstract}
The zero-temperature phases of a generalized two-leg spin
ladder with four-spin exchanges are discussed by 
means of a low-energy field theory
approach starting from an SU(4) quantum critical point.
The latter fixed point is shown to be a rich multicritical point 
which unifies different competing dimerized orders
and a scalar chirality phase which breaks spontaneously
the time-reversal symmetry.
The quantum phase transition between
these phases is governed by spin-singlet fluctuations 
and belongs to the Luttinger universality 
class due to the existence of an exact U(1) self-duality
symmetry. 
\end{abstract}            
\maketitle

Multi-spin exchange interactions have attracted much interest
over the years both theoretically and experimentally 
\cite{roger}.
Recently, ring exchange interactions have been invoked 
for the description of magnetic properties
of spin ladder compound La$_x$Ca$_{14-x}$Cu$_{24}$O$_{41}$
\cite{matsuda} and for their ability to induce new
exotic phases in quantum magnetism \cite{fisher}. 
In this respect, a scalar chirality phase \cite{wen89},
which breaks spontaneously the time-reversal symmetry,
has been found in the 
two-leg spin-1/2 ladder for a sufficiently strong
four-spin cyclic exchange \cite{lauchli03,hikihara03}.
Such an exotic ground state is, in fact, not specific to
this spin ladder and exact ground states with
scalar chirality long-range order have 
been obtained for a wider class of two-leg spin ladders 
with four-spin interactions \cite{momoi03}.
A central question is the determination 
of all possible ground states stabilized
by four-spin exchanges and the elucidation of the nature
of the quantum phase transition between these
phases.
In this letter, we will study 
a general two-leg spin-1/2 ladder
with four-spin exchanges defined by \cite{momoi03}
\begin{eqnarray}
{\cal H}_{\rm gen} &=& 
J_1 \sum_{n} \sum_{p=1}^{2}
{\bf s}_{p,n} \cdot {\bf s}_{p,n+1}
+ J_{\perp} \sum_{n}
{\bf s}_{1,n}\cdot{\bf s}_{2,n}
\nonumber \\
&+& J_d \sum_{n} \left(
{\bf s}_{1,n} \cdot {\bf s}_{2,n+1}
+ {\bf s}_{1,n+1} \cdot {\bf s}_{2,n} \right)
\nonumber \\
&+& J_{rr} \sum_{n} \left(
{\bf s}_{1,n} \cdot {\bf s}_{2,n}\right)
\left({\bf s}_{1,n+1} \cdot {\bf s}_{2,n+1}\right)
\nonumber \\
&+& J_{11} \sum_{n} \left(
{\bf s}_{1,n} \cdot {\bf s}_{1,n+1}\right)
\left({\bf s}_{2,n} \cdot {\bf s}_{2,n+1}\right)
\nonumber \\
&+& J_{dd} \sum_{n} \left( {\bf s}_{1,n} \cdot {\bf s}_{2,n+1}
\right) \left( {\bf s}_{1,n+1} \cdot {\bf s}_{2,n}\right),
\label{hamgen}
\end{eqnarray}
where ${\bf s}_{p,n}$ are spin-1/2 operators at
site $n$ on chain $p=1,2$.
The Hamiltonian (\ref{hamgen}) is the most general
translation-invariant two-leg spin ladder
which (i) consists of SU(2)-symmetric interactions 
involving two neighboring
rungs and (ii) has $\mathbb{Z}_2$ invariance under the permutation of 
the two chains ${\cal P}_{12}$.
In the following, our strategy 
is to start from a point with the maximal symmetry in a problem
consisting of two spin-1/2 operators, 
i.e., an SU(4) symmetry. 
The resulting SU(4) model displays a quantum
critical behavior which enables us to
develop a low-energy approach to investigate 
the different $T=0$ gapped phases induced 
by the SU(2)$\times \mathbb{Z}_2$-invariant symmetry breaking terms
of Eq. (\ref{hamgen}). 
As will be seen, the SU(4) symmetric point
is a rich multicritical point which unifies several 
emerging quantum orders.
The nature of the quantum phase transition between
these phases can then be determined within our approach
and belongs to the Luttinger universality
class \cite{haldane} as the result of
an exact U(1) self-duality symmetry at the
transition \cite{momoi03}.

{\it The SU(4) quantum critical point.}
The starting point of our approach is
the existence of an SU(4) symmetric point
in Eq. (\ref{hamgen}) which is
obtained for $J_{11} = 4 J_1$
and $J_{\perp}= J_d = J_{rr} = J_{dd} = 0$ \cite{li}.
The resulting model can also be regarded as the SU(4) Heisenberg 
spin chain when the four states on a rung are identified with those 
of the fundamental representation {\bf 4} of SU(4). 
The latter model is Bethe-ansatz solvable
\cite{sutherland} and displays an extended quantum
criticality, 
which is characterized by $\text{SU(4)}_{1}$ conformal field theory 
(CFT) with central charge $c=3$ \cite{affleck}.   
A simple description of this fixed point
is provided by the conformal embedding
SU(4)$_1$ $\sim$ SU(2)$_2$ $\times$ SU(2)$_2$ 
with two SU(2)s corresponding to independent rotations for 
the two chains. 
Since a single $\text{SU(2)}_{2}$ CFT is described by a triplet of 
real (Majorana) fermions, 
we may describe the critical properties of SU(4)$_1$ fixed point
by two triplets of right- and left-moving Majorana fermions
$\xi^{a}_{R,L}$ and $\chi^{a}_{R,L}$ ($a=1,2,3$). 
This Majorana fermion description is extremely useful 
to understand the symmetry properties of model (\ref{hamgen})
in the close vicinity of the SU(4) point as it
has been exploited for the SU(2) $\times$ SU(2) spin-orbital
chain \cite{nersesyan,azaria}.  
Moreover, the lattice discrete symmetries of model (\ref{hamgen}),
i.e. one-step translation symmetry (${\cal T}_{a_0}$),
time-reversal symmetry (${\cal T}$), site-parity (${\cal P}_S$), 
and the permutation ${\cal P}_{12}$ of the two chains,
are linearly represented in terms of the Majorana
fermions.
For instance, the translation symmetry is described by:
$\xi_{R}^a \rightarrow - \xi_{R}^a$,  
$\chi_{R}^a \rightarrow - \chi_{R}^a$, whereas
$\xi_{L}^a$ and $\chi_{L}^a $ are left unchanged under
${\cal T}_{a_0}$.
These results lead us to
write the most general low-energy effective
field theory for the generalized 
two-leg spin ladder (\ref{hamgen}) which is invariant
under the SU(2) spin rotational symmetry and 
the discrete symmetries ${\cal T}_{a_0} \times 
{\cal T} \times {\cal P}_S \times {\cal P}_{12}$: 
\begin{widetext}
\begin{eqnarray}
&{\cal H}& = 
{\cal H}_0
+ (g_1+g_2) \left[ \left({\vec \xi}_R \cdot {\vec \xi}_L \right)^2
+ \left({\vec \chi}_R \cdot {\vec \chi}_L \right)^2 \right]
+ (g_1-g_2) \left[ \left({\vec \xi}_R \cdot {\vec \chi}_L \right)^2
+ \left({\vec \chi}_R \cdot {\vec \xi}_L \right)^2 \right]
+ 2(g_3+g_4) \left({\vec \xi}_R \cdot {\vec \xi}_L\right)
\left( {\vec \chi}_R \cdot {\vec \chi}_L\right)
\nonumber \\
&+& 2(-g_3+g_4) \left({\vec \xi}_R \cdot {\vec \chi}_L\right)
\left({\vec \chi}_R \cdot {\vec \xi}_L\right)
- \frac{g_5}{2} \left({\vec \xi}_R \cdot {\vec \chi}_R \right)
 \left({\vec \xi}_L \cdot {\vec \chi}_L \right)
- i \frac{g_6}{2} \left( {\vec \xi}_R \cdot {\vec \chi}_R 
+ {\vec \xi}_L \cdot {\vec \chi}_L \right),
\label{fieldth}
\end{eqnarray}
\end{widetext}
where ${\cal H}_0$ is the free-Hamiltonian 
for the Majorana fermions $\xi^a_{R,L}$ and $\chi^a_{R,L}$.
No such strongly relevant mass terms for the Majorana fermions  
as ${\vec \xi}_R \cdot {\vec \xi}_L$ or 
${\vec \chi}_R \cdot {\vec \chi}_L$ are allowed since they are 
odd under ${\cal T}_{a_0}$. 
The effective Hamiltonian (\ref{fieldth}) describes
the low-energy properties of model (\ref{hamgen})
in the vicinity of the SU(4) point.
In particular, using the continuum expressions
of the spin operators ${\bf s}_{1,2,n}$ at
the SU(4)$_1$ fixed point found in Ref. \cite{azaria},
we have obtained the following identifications: 
$g_{1,2} \simeq  (J_1 \pm J_d)/2$,
$g_{3,4} \simeq (J_{11} \pm J_{dd})/8$,
$g_5 \simeq J_{rr}$, and $g_6 \simeq  J_{\perp}$.

{\it Order parameters and duality symmetries.}
Before investigating the infrared (IR) phases
of the low-energy effective field theory (\ref{fieldth}),
let us first discuss its symmetries and possible orders. 
The SU(4)$_1$ fixed point Hamiltonian, i.e. ${\cal H}_0$,
is invariant under chiral SO(2) rotations ${\cal R}_r(\theta), r =R,L$ 
on the Majorana fermions:
\begin{align}
& \xi_r^a \rightarrow \xi_r^a \cos \theta/2 - \chi_r^a \sin \theta/2 
\notag \\
& \chi_r^a \rightarrow \xi^a_r \sin \theta/2  
+ \chi^a_r \cos \theta/2 \; .
\label{eqn:so2}
\end{align}
This rotation defines a first U(1) symmetry 
${\cal U}_{\theta} = {\cal R}_L(\theta) \times
{\cal R}_R(\theta)$ which acts on the fields of
the SU(4)$_1$ CFT.
Now we introduce a first set of order parameters--%
the staggered dimerization operator
${\cal O}_{\text{SD}} = (-1)^n \left({\bf s}_{1,n} \cdot {\bf s}_{1,n+1}
- {\bf s}_{2,n} \cdot {\bf s}_{2,n+1} \right)$ 
and the scalar chirality order parameter 
\cite{lauchli03,hikihara03,momoi03}:
${\cal O}_{\text{SC}} = (-1)^n \left[ \left({\bf s}_{1,n}
+ {\bf s}_{2,n} \right) \cdot \left({\bf s}_{1,n+1}
\wedge {\bf s}_{2,n+1} \right) + (n \leftrightarrow n+1) \right]$. 
They have a simple continuum description in terms
of the Majorana fermions:
${\cal O}_{\text{SD}} \sim i \left({\vec \xi}_R \cdot {\vec \xi}_L
- {\vec \chi}_R \cdot {\vec \chi}_L \right)$
and ${\cal O}_{\text{SC}} \sim i \left({\vec \xi}_R \cdot {\vec \chi}_L
+ {\vec \chi}_R \cdot {\vec \xi}_L \right)$ and 
from Eq. (\ref{eqn:so2}) we deduce that these order 
parameters transform as a doublet under ${\cal U}_{\theta}$:
\begin{equation}
\left(\begin{array}{cccccc}
  {\cal O}_{\text{SD}}  \\
  {\cal O}_{\text{SC}}
\end{array}\right)
\rightarrow
\left(\begin{array}{cccccc}
\cos \theta  & - \sin \theta\\
\sin \theta  & \cos \theta
\end{array}\right)
 \left(\begin{array}{cccccc}
  {\cal O}_{\text{SD}}  \\
  {\cal O}_{\text{SC}}
\end{array}\right).
\label{doublet}
\end{equation}
In particular, for $\theta =\pi/2$, 
the two phases characterized by ${\cal O}_{\text{SD}}$ 
and ${\cal O}_{\text{SC}}$ 
are interchanged under ${\cal D} = {\cal U}_{\pi/2}$
which can thus be viewed as a $\mathbb{Z}_2$ duality for a pair of order 
parameters. 
Remarkably, the duality symmetry ${\cal D}$ 
and the U(1) ${\cal U}_{\theta}$
transformation have a lattice interpretation which 
has been discovered previously 
and called spin-chirality symmetries by Momoi {\it et al.} \cite{momoi03}:
${\cal U}^{\text{lat}}_{\theta} 
= \exp [-i \theta \sum_n\left({\bf s}_{1,n}
\cdot {\bf s}_{2,n} - 1/4\right)]$ and ${\cal D} = 
{\cal U}^{\text{lat}}_{\theta = \pi/2}$. 

Moreover, we can define two additional order parameters,
expressed again as bilinears of Majorana fermions: 
${\cal O}_{\text{D}} \sim i \left({\vec \xi}_R \cdot {\vec \xi}_L
+ {\vec \chi}_R \cdot {\vec \chi}_L \right)$
and ${\cal O}_{\text{RD}} \sim i \left({\vec \xi}_R \cdot {\vec \chi}_L
- {\vec \chi}_R \cdot {\vec \xi}_L \right)$
which are left invariant under the spin-chirality rotation 
${\cal U}_{\theta}$ and are thus self-dual under ${\cal D}$.
In fact, these order parameters are the continuum representation
of the columnar dimerization operator
${\cal O}_{\text{D}} = (-1)^n \left({\bf s}_{1,n} \cdot {\bf s}_{1,n+1}
+ {\bf s}_{2,n} \cdot {\bf s}_{2,n+1} \right)$
and the rung dimerization operator 
${\cal O}_{\text{RD}} = (-1)^n {\bf s}_{1,n} \cdot {\bf s}_{2,n}$.
The latter order parameter describes a phase,
with alternation of rung singlets and rung triplets,
which has been found in some integrable
two-leg spin ladder \cite{wang}. 
The second set of order parameters is closely related, 
in the continuum limit, to the existence of a second U(1) symmetry: 
${\tilde {\cal U}}_{\theta} = {\cal R}_L(\theta) \times
{\cal R}_R(-\theta)$.   
It leaves invariant the ${\cal O}_{\text{SD}}$ and 
${\cal O}_{\text{SC}}$ order parameters
whereas ${\cal O}_{\text{D}}$ and  ${\cal O}_{\text{RD}}$ 
transform now as a doublet under ${\tilde {\cal U}}_{\theta}$
as in Eq. (\ref{doublet}).
A second $\mathbb{Z}_2$ duality, 
${\tilde {\cal D}} = {\tilde {\cal U}}_{\pi/2}$,
can thus be considered as a transformation which maps 
the columnar dimerization onto the rung dimerization whereas
the staggered dimerization and the scalar chirality are
kept intact under ${\tilde {\cal D}}$. 
The SU(4)$_1$ fixed point is therefore a rich multicritical point
which unifies four different competing orders 
(dimerized ${\cal O}_{\text{D}}, {\cal O}_{\text{SD}}, 
{\cal O}_{\text{RD}}$  
and ${\cal T}$-breaking ${\cal O}_{\text{SC}}$. See Fig.1).
The SU(4) $\rightarrow$ SU(2)$\times \mathbb{Z}_2$ symmetry 
breaking perturbations
will select one of these quantum orders as we are going to
see now. 
\begin{figure}[H]
\begin{center}
\includegraphics[scale=0.4]{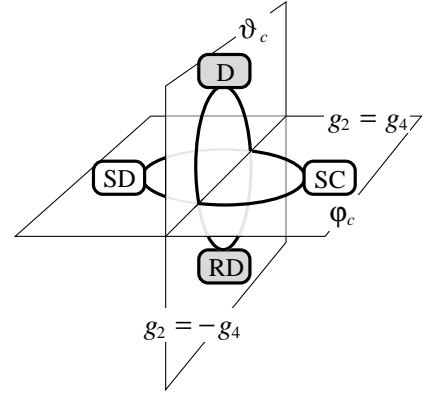}
\end{center}
\caption{Relationship between four orders. 
Note that on manifolds $g_2=g_4$ ($g_2=-g_4$) the system acquires  
U(1) symmetry under ${\cal U}_{\theta}$ ($\tilde{\cal U}_{\theta}$) 
and order parameters form a doublet. 
Two angular fields $\varphi_{\text{c}}$ and  
$\vartheta_{\text{c}}$ characterize fluctuations within each doublet 
(see the text).} 
\end{figure}

{\it Renormalization Group (RG) analysis.}
The next step of the approach is to perform 
a one-loop RG calculation to determine the nature
of the IR phases of the low-energy effective Hamiltonian (\ref{fieldth}).
First of all, 
the SU(4) model in Eq. (\ref{hamgen}), 
perturbed by a standard rung interaction $J_{\perp} \ne 0$ 
is Bethe-ansatz integrable \cite{wangbis}; 
for a small value of $J_{\perp}$, the gapless
behavior of the SU(4) model with central charge $c=3$ 
extends up to a critical point $J_{\perp c} = 4 J_{1}$ 
above which the standard gapped rung-singlet phase of the two-leg spin ladder
appears. In the close vicinity of the SU(4)$_1$ quantum critical point, 
when $|\lambda_i| \ll 1$, we can thus forget 
the perturbation with coupling constant $g_6$ in 
Eq. (\ref{fieldth}).  
Thus we are left with marginal interactions 
and the one-loop RG equations read as follows: 
\begin{eqnarray}
& &\dot{g}_{1} = g_{1}^{2}+g_{2}^{2}+5\,g_{3}^{2}+g_{4}^{2}  \notag \\
& &\dot{g}_{2} = 2\,g_{1}g_{2}+6\,g_{3}g_{4}+g_{4}g_{5} \qquad \qquad
\notag  \\
& &\dot{g}_{3} = 6\,g_{1}g_{3}+2\,g_{2}g_{4} \notag \\ 
& &\dot{g}_{4} = 2\,g_{1}g_{4}+6\,g_{2}g_{3}+g_{2}g_{5} \notag \\
& &\dot{g}_{5} = -16(g_{1}g_{3}-g_{2}g_{4})   \; . 
\label{rgeq}
\end{eqnarray}
We are now going to investigate
the different gapped phases that emerge in the IR limit
of the RG equations (\ref{rgeq}).
First of all, it is important to note that the interaction part 
of the Hamiltonian with $g_{5,6} = 0$ can be recasted as 
\begin{equation}
{\cal H}_{\text{int}} = -\lambda_{\text{SD}}{\cal O}_{\text{SD}}^{2}
-\lambda_{\text{SC}}{\cal O}_{\text{SC}}^{2}
-\lambda_{\text{D}}{\cal O}_{\text{D}}^{2}
-\lambda_{\text{RD}}{\cal O}_{\text{RD}}^{2} \; ,
\label{compet}
\end{equation}
where the couplings $\lambda_{\text{SD}}$ etc. are functions of 
$g_{1},\ldots,g_{4}$  and Eq. (\ref{compet}) describes the
competition between the quantum orders introduced previously. 
Then we apply an ansatz, proposed by
Lin {\it et al.} \cite{lin} in the context
of the half-filled two-leg Hubbard ladder, 
that the IR asymptotics of Eq. (\ref{rgeq})
is described by: $g_i(t) = r_i/(t_0 -t)$, 
where $t$ is the RG time and $t_0$ marks the crossover
point where the weak-coupling perturbation breaks down.
The coefficients $r_i$ indicate the symmetric
rays which attract the RG flow in the IR limit
and define the different strong coupling phases
of the problem. 
A first symmetric ray, $(r_1,r_2,r_3,r_4,r_5)=(1/8,1/8,-1/8,-1/8,0)$,
is described by the low-energy effective field theory \cite{azaria}: 
\begin{eqnarray}
{\cal H}_{\rm eff}^{(1)} =
{\cal H}_0
+ \lambda^{*}_{\text{SD}} 
\left({\vec \xi}_R \cdot {\vec \xi}_L 
- {\vec \chi}_R \cdot {\vec \chi}_L \right)^2,
\label{GN}
\end{eqnarray}
with $\lambda_{\text{SD}}^{*} > 0$ so that, after  
a transformation ${\vec \chi}_R \rightarrow - {\vec \chi}_R$,
the Hamiltonian (\ref{GN}) takes the form of a SO(6) Gross-Neveu (GN) 
model which is an integrable field theory 
with a spectral gap \cite{zamolo}.
The nature of the resulting strong-coupling phase can be elucidated
by observing that the interacting part of Eq. (\ref{GN})
is simply: ${\cal H}_{\rm int} = - \lambda^{*}_{\text{SD}} 
{\cal O}_{\text{SD}}^2$,
so that $<{\cal O}_{\text{SD}}> = \pm \Delta_0 \ne 0$
in the IR limit.
A staggered dimerized phase, as in the 
SU(2) $\times$ SU(2) spin-orbital chain \cite{nersesyan,azaria}, 
is thus stabilized 
and breaks spontaneously the translation symmetry.
A second symmetric ray,
$(r_1,r_2,r_3,r_4,r_5)=(1/8,-1/8,-1/8,1/8,0)$ 
is described by
\begin{eqnarray}
{\cal H}_{\rm eff}^{(2)} =
{\cal H}_0
+ \lambda^{*}_{\text{SC}} \left({\vec \xi}_R \cdot {\vec \chi}_L
+ {\vec \chi}_R \cdot {\vec \xi}_L \right)^2.
\label{GNbis}
\end{eqnarray}
Upon the chiral transformation ${\vec \chi}_R \leftrightarrow {\vec \xi}_R$,
the Hamiltonian (\ref{GNbis}) is mapped onto
the SO(6) GN model. 
The interacting part of Eq. (\ref{GNbis}) 
is written now as 
${\cal H}_{\rm int} = - \lambda^{*}_{\text{SC}}{\cal O}_{\text{SC}}^2$, 
and thus implies $<{\cal O}_{\text{SC}}> = \pm \Delta_0 \ne 0$, 
i.e. the emergence of the scalar chirality phase.  
which breaks spontaneously the time-reversal symmetry.
At this point, it is worth observing that
the Hamiltonian (\ref{GN}) and (\ref{GNbis}) are
interchanged under the duality symmetry ${\cal D} = 
{\cal U}_{\theta= \pi/2}$ on the Majorana fermions. 
In this respect, the approach used here gives the field-theoretical 
interpretation to the spin-chirality duality and the appearance of 
the long-range (staggered) scalar chiral order 
in the phase diagram of Eq. (\ref{hamgen}) which was first pointed out  
in Ref. \cite{momoi03}. 

Finally, in a similar way, there are two more 
symmetric rays where the SO(6) symmetry is restored in the 
IR limit: $(r_1,r_2,r_3,r_4,r_5)=(1/8,1/8,1/8,1/8,0)$ and  
$(r_1,r_2,r_3,r_4,r_5)=(1/8,-1/8,1/8,-1/8,0)$. 
They correspond respectively to the stabilization 
of the columnar dimerization (${\cal O}_{\text{D}}$) 
and the rung dimerization (${\cal O}_{\text{RD}}$) which
are interchanged now under the second duality ${\tilde {\cal D}}$.
In summary, 
the four phases, ${\cal O}_{\text{SD}}, {\cal O}_{\text{SC}}, 
{\cal O}_{\text{D}},{\cal O}_{\text{RD}}$, 
related two by two through the duality 
symmetries ${\cal D}$ and ${\tilde {\cal D}}$,
are the different gapped phases of the problem which 
are characterized by an SO(6) symmetry restoration
in the IR limit.

{\it Quantum phase transition.}
In addition to these SO(6) symmetric rays, there are
special manifolds where the RG equations (\ref{rgeq})
display also an larger symmetric behavior. 
On the two manifolds $g_2 = \pm g_4$, the SU(2)$\times \mathbb{Z}_2$ 
symmetric model (\ref{fieldth}) acquires a larger continuous symmetry 
SU(2) $\times$ U(1), 
being invariant under arbitrary rotations ${\cal U}_{\theta}$
and ${\tilde {\cal U}}_{\theta}$ respectively (see Fig.1). 
Within these self-dual manifolds, 
the RG flow is attracted in the IR limit towards 
two different asymptotes: 
$(r_1,r_2,r_3,r_4,r_5)=(1/6,0,\mp 1/6,0,\pm 4/9)$.  
Along the first line ($r_3=-r_1$), the interacting part of the 
low-energy self-dual theory takes the form:
\begin{eqnarray}
{\cal H}^{\rm int}_{\text{eff}} = 
-\lambda^{*} \left({\cal O}_{\text{SD}}^2 
+ {\cal O}_{\text{SC}}^2 \right) - \frac{4 \lambda^{*}}{3}
\left({\vec \xi}_R \cdot {\vec \chi}_R \right) 
\left({\vec \xi}_L \cdot {\vec \chi}_L \right),
\label{ftsdual}
\end{eqnarray}
which describes the competition between the staggered 
dimerization and scalar chirality orders, i.e.
governs the nature of the quantum phase transition between these
two phases. 
Similarly, the second asymptote accounts for
the competition between the columnar dimerization and
the rung dimerization.  
The emerging effective field theory (\ref{ftsdual}) displays a larger
symmetry than the U(1) $\times$ SU(2) symmetry 
of the initial manifold $g_2=g_4$.   
On the one hand, model (\ref{ftsdual})
turns out to be invariant not only under ${\tilde {\cal U}}_{\theta}$ 
but also under a larger ${\cal U}_{\theta} \times
{\tilde {\cal U}}_{\theta}$ symmetry 
(Note that at the lattice level, $\tilde{{\cal U}}_{\theta}$ 
may be broken by umklapp interactions).
On the other hand, it is also invariant under a hidden SU(3) symmetry.
A way to reveal this last symmetry 
is to combine the six Majorana fermions into three Dirac
fermions: $\Psi_{aR,L} = (\xi^a_{R,L} + i \chi^a_{R,L})/\sqrt{2}$.
The ${\cal U}_{\theta}$ and
${\tilde {\cal U}}_{\theta}$ rotations 
then acquire a simple meaning in terms of these Dirac fermions
since they act on them as charge U(1) chiral symmetries:
$\Psi_{a R,L} \rightarrow e^{i \theta/2} \Psi_{a R,L}$
for ${\cal U}_{\theta}$ and
$\Psi_{a R,L} \rightarrow e^{\pm i \theta/2} \Psi_{a R,L}$
for ${\tilde {\cal U}}_{\theta}$.
The self-duality symmetry ${\cal U}_{\theta} \times
{\tilde {\cal U}}_{\theta}$ of Eq. (\ref{ftsdual}) 
is described thus as an 
U(1)$_{\text{R}}$ $\times$ U(1)$_{\text{L}}$ symmetry
on the Dirac fermions. 
In addition, we can see that the order parameters ${\cal O}_{\text{SD}}$
and ${\cal O}_{\text{SC}}$ have also a simple interpretation 
as Cooper pairs by taking a combination
${\cal O}_{\text{SD}} + i {\cal O}_{\text{SC}} 
\sim {\vec \Psi}_{R} \cdot {\vec \Psi}_{L}$.
The role of these pseudo-charge
degrees freedom, introduced by the spin-chirality U(1) symmetry, 
becomes manifest with the help of a bosonization of the Dirac fermions.
To this end, we define three right-left moving 
bosonic fields $\varphi_{a R,L}$ such as:
$\Psi_{a R,L} \sim \exp[\pm i \sqrt{4 \pi} \varphi_{a R,L}]$,
and switch to a basis where the pseudo-charge
degrees of freedom single out:
\begin{eqnarray}
\varphi_{c R,L} &=& 
\frac{1}{\sqrt{3}}\left( \varphi_{1 R,L} +
\varphi_{2 R,L} + \varphi_{3 R,L}\right)
\nonumber \\
\varphi_{s R,L} &=& \frac{1}{\sqrt{2}}\left( \varphi_{1 R,L} -
\varphi_{2 R,L} \right)
\nonumber \\
\varphi_{f R,L} &=& \frac{1}{\sqrt{6}}\left( \varphi_{1 R,L} +
\varphi_{2 R,L} - 2 \varphi_{3 R,L}\right).
\label{basis}
\end{eqnarray}
In terms of these new fields, the low-energy Hamiltonian
(\ref{ftsdual}) exhibits a ``spin''(SU(3))-``charge''(U(1)) separation 
${\cal H}_{\text{eff}} = {\cal H}_{\text{c}} + {\cal H}_{\text{s}}$,
with $[{\cal H}_{\text{c}}, {\cal H}_{\text{s}}] = 0$.
The charge degrees of freedom are described by the
Tomonaga-Luttinger Hamiltonian:
\begin{equation}
{\cal H}_{\text{c}} 
= \frac{v}{2} \left[\left(\partial_x \varphi_{\text{c}} \right)^2
+ \left(\partial_x \vartheta_{\text{c}} \right)^2 \right],
\label{bosham}
\end{equation}
where $\varphi_{\text{c}} = \varphi_{\text{c} R} 
+ \varphi_{\text{c} L}$
is the total charge bosonic field and $\vartheta_{\text{c}}$ is
its dual field.
The Hamiltonian ${\cal H}_s$
for the remaining degrees of freedom
can be recasted in a fully SU(3) symmetric form
in terms of the chiral SU(3)$_1$ currents ${\cal J}^A_{R,L}$ defined from the
two bosonic fields $\varphi_{s,f R,L}$:
\begin{equation}
{\cal H}_s =
\frac{\pi v}{2} \sum_{A=1}^{8} \left({\cal J}^A_{R} {\cal J}^A_{R}
+ {\cal J}^A_{L} {\cal J}^A_{L} \right)
+ g^{*} \sum_{A=1}^{8} {\cal J}^A_{R} {\cal J}^A_{L}.
\label{GNsu3}
\end{equation}
The latter model is the SU(3) GN model which 
is a massive ($g^{*} >0$) integrable field theory \cite{andrei}.
A spectral gap is thus formed by the interactions
in the ``spin'' sector and the low-lying excitations
are known, from the exact solution \cite{andrei}, 
to be massive SU(3) spinons and antispinons.  
The low-energy physics of (\ref{ftsdual}) is dominated by 
the gapless spin-singlet fluctuations (\ref{bosham}) 
of the ``charge'' degrees of freedom which stems from
the remarkable U(1)-symmetry (${\cal U}_{\theta}$) 
of Eq. (\ref{ftsdual}).  
Therefore we may conclude that 
the quantum phase transition between the staggered dimerized- 
and scalar chirality phases belongs to the $c=1$ 
Luttinger-liquid universality class.  
The physical properties at the transition can also
be determined within our approach. 
At the transition, all order parameters have zero
expectation values: $\langle {\cal O}_{\text{SD}} \rangle = 
\langle {\cal O}_{\text{SC}} \rangle = \langle {\cal O}_{\text{D}} \rangle 
= \langle {\cal O}_{\text{RD}} \rangle = 0$. 
The first doublet ${\cal O}_{\text{SD}}$ and ${\cal O}_{\text{SC}}$
has a fixed modulus and correlation functions decaying 
as $x^{-2/3}$, i.e. has long-range coherence,  
whereas the second one ${\cal O}_{\text{D}}$ and ${\cal O}_{\text{RD}}$ 
is exponentially decaying due to strong quantum fluctuations. 
Now it is straightforward to discuss the effect of 
a small deviation from the self-dual manifold (\ref{ftsdual}) 
by switching on the perturbation:
${\cal V} = \epsilon \left({\cal O}_{\text{SD}}^2 - 
{\cal O}_{\text{SC}}^2 \right), |\epsilon| \ll 1$
which breaks in particular the ${\cal U}_{\theta}$ symmetry of
model (\ref{ftsdual}).
This small symmetry-breaking perturbation does not
close the spin gap but the charge Hamiltonian
(\ref{bosham}) acquire now 
a `pinning' term ${\cal V}_c \simeq 
- \epsilon \cos (\sqrt{16 \pi/3} \; \varphi_c)$ 
and becomes equivalent to a quantum sine-Gordon model. 
The interaction has scaling dimension $\Delta = 4/3 < 2$
so that the perturbation opens a charge gap; 
for $\epsilon <0$ (respectively $\epsilon  > 0$), 
the staggered dimerization (respectively scalar chirality)
order is stabilized by the small symmetry-breaking term.
The same argument applies to the second ray ($r_1=r_3$) as well 
after the replacement $\varphi_{c} \leftrightarrow \vartheta_{c}$ 
and describes the competition between ${\cal O}_{\text{D}}$ and 
${\cal O}_{\text{RD}}$.  

In summary, we have shown, in the continuum approach, that   
four different gapped phases around the SU(4) multicritical point 
are unified by the hidden 
$\mathbb{Z}_2$ symmetries ${\cal D}$ and $\tilde{\cal D}$.  
The spin-chirality U(1) ${\cal U}_{\theta}$ symmetry plays an essential role 
on the self-dual manifold and as a consequence 
a second-order phase transition which separates 
the staggered dimerized- and scalar chirality phases
is characterized by a $c=1$ Luttinger-liquid fixed point.   
This point was unclear in the preceding papers mainly because 
of numerical limitations.  
On the basis of this fact, we explained 
how an exotic phase with a broken time-reversal symmetry 
is stabilized.   
Moreover, by a mapping onto a low-energy effective theory, 
we have revealed another hidden relationship 
between columnar-dimer phase and rung-dimer phase together with 
a corresponding U(1) symmetry ${\tilde {\cal U}}_{\theta}$. 
Finally, a similar unifying approach based on an U(4) symmetry
can also be performed to describe the exotic 
phases and transitions in doped generalized two-leg
ladders as it will be discussed elsewhere.

The authors are very grateful to E. Boulat 
for illuminating discussions.
They would like also to thank P. Azaria, A.
L{\"a}uchli, T. Momoi, E. Orignac for very useful discussions.


\begin{thebibliography}{101}

\bibitem{roger}
D. J. Thouless, Proc. Phys. Soc. London {\bf 86}, 893 (1965);
M. Roger, J. H. Hetherington, and J. M. Delrieu, 
Rev. Mod. Phys. {\bf 55},  1 (1983).
\bibitem{matsuda}
M. Matsuda {\it et al.}, Phys. Rev. B {\bf 62}, 8903 (2000).
\bibitem{fisher}
A. Paramekanti, L. Balents, and M. P. A. Fisher, Phys. Rev. B {\bf 66}, 
054526 (2002).
\bibitem{wen89}
X. G.~Wen, F.~Wilczek, and A.~Zee, Phys.Rev. B {\bf 39}, 11413 (1989).
\bibitem{lauchli03}
A. L{\"a}uchli, G. Schmid, and M. Troyer, Phys. Rev. B {\bf 67},
100409(R) (2003).
\bibitem{hikihara03}
T. Hikihara, T. Momoi, and X. Hu, Phys. Rev. Lett. {\bf 90},
087204  (2003).
\bibitem{momoi03}
T. Momoi, T. Hikihara, M. Nakamura,
and X. Hu, Phys. Rev. B {\bf 67}, 174410 (2003).
\bibitem{haldane}
F. D. M. Haldane, Phys. Rev. Lett. {\bf 45}, 1358 (1980).
\bibitem{li}
Y. Q. Li {\it et al.},
Phys. Rev. Lett. {\bf 81}, 3527 (1998);
Y. Yamashita, N. Shibata, and K. Ueda,
Phys. Rev. B {\bf 58}, 9114 (1998).
\bibitem{sutherland}
B. Sutherland, Phys. Rev. B {\bf 12}, 3795 (1975).
\bibitem{affleck}
I. Affleck, Nucl. Phys. B {\bf 305}, 582 (1988).
\bibitem{nersesyan} A. A. Nersesyan and A. M. Tsvelik,
Phys. Rev. Lett. {\bf 78}, 3939 (1997). 
\bibitem{azaria}
P. Azaria {\it et al.},
Phys. Rev. Lett. {\bf 83}, 624 (1999);
P. Azaria, E. Boulat, and P. Lecheminant, 
Phys. Rev. B {\bf 61}, 12 112 (2000).
\bibitem{wang}
Y. Wang, Int. J. Mod. Phys. B {\bf 13}, 3323 (1999);
G. Albertini, Phys. Rev. B {\bf 64}, 094416 (2001). 
\bibitem{wangbis}
Y. Wang, Phys. Rev. B {\bf 60}, 9236 (1999).
\bibitem{lin}
H.-H. Lin, L. Balents, and M. P. A. Fisher,
Phys. Rev. B {\bf 58}, 1794 (1998).
\bibitem{zamolo}
A. A. Zamolodchikov and A. B. Zamolodchikov, 
Ann. Phys. (N.Y.) {\bf 120}, 253 (1979).
\bibitem{andrei}
N. Andrei and J. H. Lowenstein, Phys. Lett. B {\bf 90}, 106 (1980).

\end{thebibliography}
\end{document}